\documentclass[12pt]{article}

\usepackage{scicite}

\usepackage{times}
\usepackage{changes}
\usepackage{amsmath}
\usepackage{amsfonts}
\usepackage{amssymb}
\usepackage{graphicx}
\usepackage{bm} % this is us
\usepackage{color}
\usepackage{changes}
\usepackage[normalem]{ulem}

\usepackage{totcount}
	\regtotcounter{table} 	%count tables
	\regtotcounter{figure} 	%count figures

\newcommand\wordcount{
    \immediate\write18{texcount -sum -1 \jobname.tex > count.txt} \input{count.txt} }

\newenvironment{sciabstract}{%
\begin{quote} \bf}
{\end{quote}}

\title{Planning for Electric Vehicles Coupled with Urban Mobility}

\author
{Yanyan Xu,${}^{1\dag}$ Serdar \c{C}olak,${}^{1,2\dag}$ Emre C. Kara,${}^{3}$ Scott J. Moura,${}^{4}$ \\Marta C. Gonz\'{a}lez${}^{1,5\ast}$ \\
\\
\normalsize{${}^{1}$Department of Civil \& Environmental Engineering,}\\
\normalsize{MIT, Cambridge, MA, 02139, USA}\\
\normalsize{${}^{2}$Lawrence Berkeley National Laboratory, Berkeley, CA, 94720, USA}\\
\normalsize{${}^{3}$SLAC National Accelerator Laboratory, Menlo Park, CA, 94025, USA}\\
\normalsize{${}^{4}$Department of Civil \& Environmental Engineering,}\\
\normalsize{UC Berkeley, CA, 94720, USA}\\
\normalsize{${}^{5}$Center for Advanced Urbanism, MIT, Cambridge, MA, 02139, USA}\\
\\
\normalsize{$^\dag$Equal contribution}\\
\normalsize{$^\ast$To whom correspondence should be addressed; E-mail:  martag@mit.edu.}
}

\date{}

\begin{document} 


\baselineskip24pt


\maketitle

%{\color{red}\textbf{The total number of words: \wordcount}}

%\linenumbers

\begin{sciabstract}
The rising adoption of plug-in electric vehicles (PEVs) leads to the alignment of their electricity and their mobility demands. Therefore, transportation and power infrastructures are becoming increasingly interdependent. In this work, we uncover patterns of PEV mobility by integrating for the first time two unique data sets: (i) mobile phone activity of 1.39 million Bay Area residents and (ii) charging activity of PEVs in 580,000 sessions obtained in the same region. We present a method to estimate individual mobility of PEV drivers at fine temporal and spatial resolution integrating survey data with mobile phone data and income information obtained from census. Thereupon, we recommend changes in PEVs charging times of commuters at their work stations that take into account individual travel needs and shave the pronounced peak in power demand. Informed by the tariff of electricity, we calculate the monetary gains to incentivize the adoption of the recommendations. These results open avenues for planning for the future of coupled transportation and electricity needs using personalized data.
\end{sciabstract}

%======================================
\section*{Introduction}
The excessive exploitation of petroleum and coal affect not only the security of energy supply but also air quality and climate change. These shortcomings have triggered the search for cleaner alternative fuels for transportation~\cite{michalek2011valuation,atia2015more,needell2016potential}.
Today's plug-in electric vehicle (PEV) technology is one of the most promising candidates up to date~\cite{nykvist2015rapidly,melton2016moving}. The main issues that hindered their adoption were: range anxiety, charger unavailability, and high prices~\cite{needell2016potential}. However, improvements in battery technology, tax breaks and subsidized charging programs~\cite{hu2016integrated, deshazo2016improving} have improved these limitations. As a result, PEVs are becoming a more viable means to move and are being adopted by drivers at steadily increasing rates~\cite{nykvist2015rapidly}. According to the US Energy Information Administration, the number of PEVs in the USA doubled between 2013 and 2015 and is expected to reach 20 million by 2020~\cite{ev_outlook}. 

Planning for the mobility needs of PEVs is particularly important in the context of the vulnerability of the power grid to outages that can cascade drastically~\cite{hines2009large}. Large scale failures signified a need to reexamine the balance between power demand and the electricity infrastructure, opening the need for interdisciplinary approaches to study this complex system~\cite{brummitt2013transdisciplinary}.
A body of literature has focused on the nature of network reliability of power grids, the role of network topology on the spread of cascading failures~\cite{buldyrev2010catastrophic,brummitt2012suppressing,pahwa2014abruptness,McAndrew2015,mwasilu2014electric,halu2016data,mureddu2015green}. In the subject of PEVs and their impact on the grid, methods of optimization and control of PEV electricity consumption have a rich set of avenues~\cite{bayram2012smart,callaway2011achieving,moura2011stochastic}. Research topics on this front include measuring impact on the grid~\cite{clement2010impact,tal2014charging,harris2014empirically,lin2014optimizing,rajakaruna2015plug,tamor2015rapid,hines2014understanding}, developing accurate PEV energy consumption models~\cite{yuksel2015effects}, energy management~\cite{rezaei2014packetized,valogianni2014effective}, smart charging strategies that probe centralized and decentralized approaches~\cite{ma2013decentralized,Kara2015515}, scheduling~\cite{subramanian2013real,yang2014risk}, peak shaving, emissions, pricing models~\cite{zakariazadeh2014multi,garcia2014plug}, and joint optimization of power and transportation networks~\cite{alizadeh2015joint}. A common shortcoming in these works is the narrow scope in incorporating individual mobility needs into the analyses, often limited to the estimation of arrival or departure hours. Up to date, data on individual mobility demand at metropolitan scale has not yet been incorporated into the planning schemes to manage electricity demand. 

In this work, we target these gaps in the literature to extend the current knowledge of transportation based electricity. For this purpose, we bring together three independent data sources: (i) mobile phone activity of a large sample of the residents of the San Francisco Bay Area, (ii) charging sessions obtained from the commercial PEV supply equipment in the same region, and (iii) surveys on the use of conventional and electric vehicles, together with census data for income information at the zipcode level (see Data in Materials and Methods). In the first part of the work, we estimate individual vehicular mobility per week day in the Bay Area using the mobile phone activity of a large sample of residents. We then present a Bayesian methodology to sample the PEV drivers from all travelers by utilizing information obtained from surveys regarding the household income and daily travel distances of PEV drivers. In the second part, with the charging session data, we analyze the various aspects of charging activity to characterize the nature of electricity demand at charging stations. We present observations regarding visitation frequencies, arrival and departure hours, typical per session energy consumption patterns, and power levels. We observe that PEV charging patterns are highly regular with morning and evening peaks following the traffic peaks. These peaks of demand are undesired because they can cause instabilities in the power grid. 
%\deleted{The opportunity is that charging sessions have temporal flexibility, that is, PEVs are typically charged for only a portion of the time they stay plugged in.} 
Currently, the power consumption happens immediately upon arrival and makes no use of this flexibility. 
To tackle this problem, we explore the relationship between the consumption of simulated PEV commuters working in the selected zipcodes and the observed energy demand at individual commercial charging stations in the same region. We calibrate the charging behaviors of PEV drivers to match the observed demand. As an application, we lay out a charging scheme that minimizes the peak power changing the start and end of the charging sessions, while also taking into account the constraints in changing departures and arrivals.
We show how not knowing the mobility constraints decreases the potential of the peak minimization schemes. In contrast, introducing the awareness of individual mobility increases the feasibility of their adoption, affecting less the benefits of peak minimization. The resulting effects on the commuting travel times and the monetary benefits from the changes in charging times support the viability of the charging time shifts. Fig. 1 depicts a summary of the proposed framework.

\section*{Results}
% =================================================
\paragraph*{Estimating Individual Mobility of PEV Drivers\label{sec:evODs}}

We simulate the individual mobility of the entire population of the Bay Area using a fine-scale urban mobility model, TimeGeo~\cite{jiang2016timegeo}. This process begins with the extraction of stay locations in the trajectories of each individual~\cite{jiang2013review,colak2014analyzing,toole2015path}. Each location is then accordingly labeled as \emph{home}, \emph{work}, or \emph{other}, based on temporal properties of the call activities. Sequences of trips are collected for each individual and categorized by their time and purpose. According to whether the workplaces are detected or not, we model the trips of commuters and non-commuters respectively (see Individual Mobility Model in Materials and Methods). Fig. 2A represents the simulated trajectory and the labeled activities of a mobile phone user. The simulations of individual mobility based on mobile phone data, compare very well with the results using two travel surveys, the 2010-2012 California Household Travel Survey (CHTS)~\cite{CHTS2010}, and the 2009 National Household Travel Survey (NHTS)~\cite{NHTS2009}. As shown in Fig. 2B-D, the daily visited locations and fraction of departures per time of the day agree well between our model based on phone data and the travel surveys. Further comparisons between our model and the travel diaries are presented in the Supplementary Figs. 1 and 2. 

Mobility motifs~\cite{schneider2013unravelling} describe the individual daily trips. Specifically, motifs constitute directed networks where nodes are visited locations and edges are trips from one location to another. For example, the motif of an individual whose only trips in a day are to and from work will consist of two nodes with a two directed edges (one in both directions). While on average, individuals visit three different places per day, when constructing all possible directed networks with $6$ or fewer nodes, there exist over $1$ million ways for an individual to travel between. However, $90\%$ of people use one of just $17$ networks, called motifs~\cite{schneider2013unravelling}. While nearly half of the population follow the simple two-locations motif. These results can be modeled with a probabilistic Markov model~\cite{jiang2016timegeo} that assigns particular rates to each individual informed by their trip behavior. The top $10$ motifs of nearly $6$ million simulated drivers in the Bay Area are summarized in Fig. 2C., the distribution of our simulated motifs agrees well with the information gathered from mobile phone users. 
After simulated individual mobility overall, we can probabilistically estimate the individual mobility of PEVs. To that end, we make use of the vehicle usage rate from the US census data and the California Plug-in Electric Vehicle Driver Survey~\cite{ev_survey}. According to this survey, PEV drivers' income distribution is skewed towards higher income segments. In particular, the percentage of those with average annual income above 150K\$ among conventional vehicle drivers is 15\%, compared to the 47\% observed among PEV drivers. The survey also highlights the typical distances PEV drivers travel: 64\% of PEV drivers travel less than 30 miles per day (see Table 1). This information is used to subsample PEV trips from total vehicular trips by implementing the Bayesian sampling procedure. Namely, we use the individual income estimated from the US census data at the census tract level and daily route distance from TimeGeo to estimate the probability of that the driver travels with a PEV, both for commuters and non-commuters (see Electric Vehicle Mobility Estimation in Materials and Methods).

Figure 3A depicts the number of PEVs estimated from the Bayesian method at each zipcode and the number obtained from the data set on PEVs collected by the California air resources board clean vehicle rebate project~\cite{CVRP}, that will be referred here as the CVRP data set. Fig 3B shows a good agreement between the number of PEVs obtained via the Bayesian estimates and the mobility model vs. the ground truth of PEVs usage. Fig. 3C-D compare the distributions of the morning route distance, $D$, made by all commuters vs. PEVs drivers, as well as the commuting travel time, $T$, under free flow conditions. There are fewer PEV trips shorter than $5$ km and longer than $25$ km, in agreement with the findings of the survey. Figure 3E depicts the four mobility motifs from PEV commuters, showing that approximately $60\%$ of PEV commuters mostly travel between home and work during weekdays. The simple motifs (with ID=1) is more prevalent among PEV drivers than among commuters using conventional vehicles, this may be a sign of the driver's concerns on the range of PEVs.

% =================================================
\paragraph*{Electric Vehicle Charging Session Data Profiles\label{sec:sessions}}
In this section, we analyze PEV charging in non-residential regions by examining: visitation patterns and adoption rates, temporal features of arrivals and departures, and typical energy and power consumption levels. PEV drivers display varying degrees of regularity in terms of how often they visit charging stations. Fig. 4A reveals that for the majority of PEV drivers the average number of charging sessions per day, $N_{day}$ is less than 1. The bottom left inset in Fig. 4A displays the logarithmic distribution of the number unique PEV charging stations (EVSEs) visited by each PEV driver, $N_{EVSE}$. Noticeably the great majority of PEV drivers (95.6\%) is observed in less than 20 distinct EVSEs. The top right inset of Fig. 4A depicts the rate of PEV adoption observed throughout the year. The 3000 drivers observed in January 2013 increases by an average of 1000 per month, doubling twice over the course of 2013. 

We look at the arrival and departure hours of charging sessions, $h_{a}$ and $h_{d}$, in Fig. 4B. Approximately 50\% of all arrivals take place in the 6am-11am morning period, and as expected, the morning and the evening peaks are highly pronounced. This points to the parallels between the temporal component of overall travel demand to electricity demand. We compare the distribution of departure time in the morning of commuters with the arrival time of charging sessions and find notable delay between these two distributions (see Supplementary Fig. 6A). Such delay represents the driving time of commuters from home to work in the Bay Area, which is around 30 min on average~\cite{Commute2017}. In the inset of Fig. 4B we look at the distribution of inter-arrival and inter-departure times, $\Delta h_{a}$ and $\Delta h_{d}$, i.e. the time between two consecutive charging sessions for the same driver ID. These distributions are peaked at multiples of 24 hours, pointing to the diurnal periodicity of PEV drivers' charging behavior. These findings reinforce the notion that commuting and charging behavior in the non-residential regions are highly related.

Next, we shift our focus to measures per session, such as energy, duration, and power. Fig. 4C exhibits the average energy consumption per session, $E_{S}$. The battery sizes of Nissan Leaf (24 kWh) and Chevrolet Volt (16 kWh), two of the most commonly used PEVs in the region are marked (see also Supplementary Fig. 4)~\cite{CVRP}. Typically $E_S$ are well below these capacities, indicating that PEV drivers typically stay within the range of their PEVs. PEV drivers can charge at home or they not necessarily start their commute at full capacity. On the other hand, the distribution of session durations reveals that 98.4\% of all charging sessions last less than a day (Fig. 4C), in line with the ubiquitous charging at work places. Given the flexibility in terms of battery capacity and mobility patterns, here we assume that the session energy $E_S$ represents not a single commuting trip, but rather a number of them.

The actual charging activity does not last as long as the session duration $\delta_S$, as seen in Fig. 4D. We divide sessions into four categories based on their session duration, and plot the average power consumption for each segment at various percentages of the total duration. We observed three levels of power rate that are most common, denoted here as Level 1 (L1), Level 2 (L2) and Level 3 (L3). The first two deliver 120V and 240V, typically corresponding to 3.3kW and 6.6kW, respectively. L3 chargers are mainly for fast charging at 480V and are relatively uncommon. As faster charging technology becomes more abundant, the peak load yielded by PEVs will be even higher as the charging sessions start intensively in the morning. In the charging dataset, L1 and L2 chargers make up 99.9\% of all the sessions in the dataset (see Supplementary Fig. 6C). This composition of power ratings explains the 4~kW upper limit to average power consumption observed in Fig. 4D. For sessions lasting less than 4 hours, average power stays above 3~kW up to 80\% of the duration into the session. Conversely, for sessions that last longer than 12 hours but less than a day, only in the starting 25\% of the session duration there is active charging. This corresponds approximately to 3-6 hours, and the power remains zero thereafter. This is consistent with constant-current constant-voltage battery charging behavior and it suggests that currently there is no strategy to charging involved: PEVs are charged immediately upon arrival.

% =================================================
\paragraph*{Coupling PEVs Energy Demand and Individual Mobility Patterns\label{sec:coupling}}
This section presents the coupling of the mobility model with the energy demand at each destination. First, we measure the distribution of electricity demand at a zipcode from the charging sessions, we connect that to the distribution of estimated energy demand of simulated PEVs commuting to that zipcode. The charging sessions data is provided by a private company with a partial coverage of the market with reasonable agreement of the most popular destinations for charging (see Data in Materials and Methods, and the comparison of estimated PEVs vs. the charging data in Supplementary Fig. 5).
To estimate the energy demand of each PEV trip, we first assign each PEV a mode from the four most popular modes (see Supplementary Fig. 4). Each PEV mode is associated an energy consumption model. Specifically, we use the drivetrain model for the two battery electric vehicle (BEV) modes, Nissan Leaf and Tesla Model S~\cite{Saxena2015265}, and the charge-depleting model for the two plug-in hybrid electric vehicle (PHEV) modes, Chevrolet Volt and Toyota Prius~\cite{wu2014cost}. For each PEV trip, we estimate the energy demand using its average speed and route distance (see details in Energy Consumption Models of Materials and Methods and in Supplementary Fig. 6B).
%\deleted{To each PEV trip, given its average speed and route distance information of the trip we assign an energy demand via a drivetrain model}~\cite{Saxena2015265}. \deleted{As a benchmark, we implement the drivetrain model with the energy per mile of different averages speeds of Nissan Leaf presented in Ref.}~\cite{needell2016potential} \deleted{(see details in Drivetrain Model of Materials and Methods and in Supplementary Fig. 6B).}

Due to the limited information from charging sessions data, we can not infer the state of charge of each vehicle. Therefore, we assign different shares of charging states: morning consumption, daily consumption, and two-days consumption. This corresponds to different charging behaviors respectively: charging both at home and work every day, charging at work once per day, or charging at work once every other day, indicating that the energy consumption at the arrival equals to the consumption of the trips in the last two days. In Fig. 3F, we show the comparison of probability distributions of the energy consumption of the three scenarios together with the actual charge, $E_{S}$ in a selected zipcode. The peaks in the distribution of $E_{S}$ demonstrate the heterogeneity in the electricity demand, which is mainly caused by the travel distance, the battery capacities, and the charging behaviors of various PEVs. The 3-4 kWh peak, which can be observed in both actual and estimated consumptions, is a combination of low energy demand as a consequence of short commuting trips and PHEVs that typically have a battery capacity around 4 kWh~\cite{yilmaz}. Further comparisons between daily consumption estimates and the charging station data in selected zipcodes are shown in Fig. 3G. The charging data have more pronounced peaks than our daily curves, this may be because our charging behaviors are simplified, leaving room for further improvements.

To estimate the charging behavior in the given zipcode, we limit the amount of charging sessions to 30 kWh. The charging behavior is distributed to match the demand of the zipcode with most charging sessions. Corresponding to average charging schemes that result in: $10\%$, $35\%$, $55\%$ of the PEVs drivers respectively (see also Supplementary Fig. 6C). Note that only $10\%$ of the drivers charging at home is in agreement with a recent report by the Department of Energy that states that 80\% of partners in their Workplace Charging Challenge program provide free PEV charging~\cite{oleksek}. We randomly assign the charging speed from the charging session data to the simulated PEVs to make the distribution of charging speed match with the ground truth (see Supplementary Fig. 6D).

%================================================
\paragraph*{Strategies to mitigate peak demand coupling energy demand with mobility needs\label{sec:optim}}
Our goal in this section is to transform the load curve into one that is more uniformly distributed across the day. To that end we propose changes in the start and end of the charging sessions such that the total load is minimized. 

We cast the problem as a mixed-integer linear program with discrete shifts in arrival times and charging end times as inputs (see Optimization Model in Materials and Methods). The program modifies the total power $P_t$ measured through the day resulting from the overlapping charging activities of a population of PEVs in a way that minimizes the peak power while keeping the total energy consumed constantly.
In this context, we test two different strategies. 
%\deleted{The first fixes the arrival times for PEVs and delays the charging by $d^{i}$, an amount specific to session $i$ within the interval $[0,d]$. The drivers are free to arrive as they wish, however, the PEV charging can be delayed to minimize $P_{peak}$. We refer to this strategy as \textit{start bound}.}
The first fixes the departure times for PEVs and shifts the arrival time in advance by $d^{i}$, an amount specific to session $i$ within the interval $[0,d]$, to minimize the peak power load $P_{peak}$. We refer to this strategy as \textit{end bound}.
The second strategy, referred to as \textit{flexible}, offers modifications to both of the arrival and departure times. In this approach, charging activity is shifted in the interval $[-d,d]$. In both of the two strategies, the PEVs are charged once they are plugged in the charging station and PEV driver could depart at their scheduled time if the charging session has finished. As a future scenario, we show the peak load saving when the charging station is able to control the PEV start of the charging session independent of the arrival time. The current infrastructure does not allow the start of charging at an optimized time and it is couple to the PEV arrival. With a smarter charging-shift scenario, the charging could be freely shifted between the plug-in/arrival and plug-out/departure time. We test the time shifting scenarios on the $448$ PEVs traveling to our zipcode with the largest number of incoming users.

Fig. 5A illustrates an instance where an actual section is of $2.25$ hours and it is modified in $d=4$ (60 minutes) to start earlier. The right panel of the figure shows all charging sessions for the \emph{flexible} strategy where some users are charged earlier and others later than their actual request.
Fig. 5B depicts how the power curves are modified under the \emph{flexible} strategy with varying value of $d$ (Supplementary Fig. 7 presents the results for the \emph{end bound} strategy). The \emph{flexible} strategy reduces $P_{peak}$ down $47\%$ from $1019$ kW to approximately $479$ kW for $d=4$, or 1 hour. In contrast, the \emph{charging-shift} strategy reduces the peak load by $66\%$ as we have more room to operate the PEV charging.

We also evaluate the effects of introducing by the constraints of the individual mobility motifs of each PEV. Mobility constraints are introduced via the $4$ motifs depicted in Fig. 3E. More than $30\%$ of PEV drivers have other activities before or after work, therefore, they are limited to accept the recommendations of a time-shift strategy if the recommendation falls before their usual arrival and departure times. Namely, we impose the following restrictions per motif ID. (1) \emph{home-work-home}, can change both of the arrival and departure time; (2) \emph{home-work-other-home}, indicates activities after work, and they can not delay their departure time; (3) \emph{home-other-work-home}, which indicates activities before work, and they can not change their arrival time; (4) \emph{home-other-work-other-home}, which indicates the PEV drivers can change neither the arrival nor the departure time.

Figure 5C contrasts the estimates of peak saving with and without the consideration of individual mobility constraints in the optimization strategy, looking at the percentage of peak loads of three schemes with variant $d$. The three schemes are (i) The \emph{Optimal} strategy, where all PEV drivers can follow the time shifts; (ii) The \emph{Motif blind} subtracts to the \emph{Optimal} estimates, all the PEV drivers that will not accept the recommendations due to the individual mobility constraints (iii) The \emph{Motif aware} is a customized strategy informed by the individual mobility constraints to distribute the time shifts. These three strategies are evaluated under the \emph{end bound} and \emph{flexible} schemes. The comparisons of the power curves of the three schemes and the two strategies are shown in Supplementary Fig. 7 and 8. 

The \emph{Optimal} results are the best case scenario because all sessions can be shifted. The \emph{Motif aware} scheme represents the more feasible gains by coupling the charging strategies with the constraints of the drivers. The \emph{Motif blind} scheme is added to show how optimization strategies based on charging data only overestimate the benefits of the savings from $3\%$ to $15\%$, while this loss can be overcome with the mobility information.

For the \emph{Motif aware} scheme, we examine the peak load saving versus the PEV driver's adoption rate (a.k.a. flexibility) of the time shifts. Fig. 5D shows a linear relationship with the acceptance rate and the savings for both of the \emph{Motif blind} and \emph{Motif aware} schemes. With the increase of adoption rate, \emph{Motif aware} contributes more on peak load saving than \emph{Motif blind}. For instance, when $80\%$ of PEV drivers accept the time-shift recommendations, the \emph{Motif aware} scheme reduces the peak load by 424kW on average, while the \emph{Motif blind} scheme reduces 279kW on average.

In Fig. 5E, we present the peak load saving of the three flexible schemes versus the number of PEVs traveling to the selected zipcode. The inset of Fig. 5E shows the estimated peak load without energy demand management. Both the peak load and the three saving powers grow linearly with the numbers of PEVs. The gap between \emph{Optimal} and \emph{Motif aware} is negligible in comparison with \emph{Optimal} vs. \emph{Motif blind}.

This framework allows us to evaluate how the time shifts in their departures affect the commuting travel times of the PEV trips into the subject zipcode. Figure 5F shows that the peak load reductions can be achieved without causing major discomfort to commuters in terms of travel times. The most negatively influenced drivers end up losing a maximum of 20 minutes in the case of $d=4$ (1 hour), and are far less than those who are unaffected by the proposed changes. There is a number of drivers that even achieve travel time savings.

Finally, we examine the monetary outcomes of the proposed strategies, \emph{end bound} and \emph{flexible}. We use the E-19 rate structure for the region to calculate the change in demand charge as a proxy of the cost in terms of dollars~\cite{billing, Kara2015515}. We use the max part-peak demand summer rates. The peaks we observe fall in the shoulder period of 8:30am-12:00pm, for which the demand charge rate is 4.07\$/kW. When implemented, the possible benefits of the schemes we proposed are displayed in Fig. 5G: monthly potential savings in the demand charge can reach up to 2500\$ for the motif customized and \emph{flexible} strategy, correspond to roughly 5\$ per month per session. Without managing charging, these savings remain unrealized, and are paid by PEV drivers or the companies that subsidize the charging activity. As a sum the savings are substantial, yet for the number of sessions on a typical weekday considered here, the amount saved per individual is relatively small, making the uniform distribution of savings a relatively unexciting reward for cooperation. However, as the biggest beneficiary of the PEV traveling management, the power grid operator could pay the PEV drivers to encourage their initiatives on the travel schedule shift. In addition, recent studies have suggested that gamified systems are successful in promoting behavior that helps achieve social good~\cite{merugu2009incentive}. More specifically, these systems encourage engagement by building raffles in which each participant has a chance to win a bigger reward with a probability proportional to their cooperation level. This type of mechanisms may make incentivization more attractive in the context of PEVs and their electricity demand management.

Moreover, we analyze the potential impact of the PEV travel demand management strategy on the non-EV charging electricity demand in both of the residential regions and commercial buildings. Among the PEV drivers involved in the time-shift strategy, $60\%$ of them are recommended to depart from home earlier and $40\%$ are recommended to depart from home later. Considering most of the household electricity usages have two peaks, morning and evening peak~\cite{Xu2017}, we argue that the time-shift strategy is more likely to relieve the morning peak of electricity usage in the residential area as a part of PEV drivers are leaving home earlier than usual. Similarly, as the PEV drivers arrive home at variant time, the time-shift strategy is also likely to relieve the evening peak load. For the commercial buildings, the power load curve is more stable during the working hours~\cite{luo2017electric}. The change of arrival time to or departure time from the commercial building will not have significant influence on the power load.

%================================================
\section*{Discussion}
This work presents, to the best of our knowledge, the first exploratory analysis that couples two unique large datasets on urban mobility and energy consumption of electric vehicles. We address a gap in existing PEV management literature, namely the simplistic modeling of urban mobility, by generating a model of individual mobility informed by large scale mobile phone data. Moreover, we extend the proposed methodology by demonstrating a charging management scheme and assess its applicability by using the information on individual mobility constraints of the drivers.

We evaluate recommendation schemes of time shifts in the charging sessions constrained by the individual mobility motif of PEV drivers. That is, about $30\%$ of PEV drivers could be limited to change their travel plan due to their schedule in other activities before or after work. Following these, peak power values can be shed by up to $47\%$. To assess the feasibility of the recommendations, we estimate the possible monetary benefits and the travel time losses resulting from the proposed schemes. The resulting daily savings while modest at the individual level, are certainly substantial enough to fund game based prizes that induce cooperation and raise awareness. On the other hand, the travel time losses are almost imperceptible to the majority of the drivers, and a substantial number of drivers in fact benefit from the adjustment of their arrival times as it aids them to escape morning traffic. 

%\deleted{The presented framework relies on individual location data from mobile phones and a survey of PEV drivers. While in the profiles of PEV adopters may change due to the rebate policy, new battery technique, etc., the surveys can be updated based on sales. Thus, the proposed model is adaptable and can be used to evaluate different scenario analysis of energy demand of PEVs in time and space.}

The presented framework relies on individual location data from mobile phones and a survey of PEV drivers. We designed a Bayesian inference framework to estimate the PEV usage probability of each vehicle driver. The Bayesian inference framework relies on three properties of the PEV users: the distribution of their household income, the distribution of their daily driving distances, and the adoption rate of PEV in the city. While in the profiles of PEV adopters may change due to the rebate policy, new battery technique, etc., the surveys can be updated based on sales. Thus, via changing one or more properties, it’s easy to estimate the use of PEV in each zipcode under different scenarios. Associated with the mobility information, the proposed model is adaptable and can be used to evaluate different scenario analysis of future energy demand of PEVs in time and space. In contrast, the prevalent data-driven energy demand methods mainly predict the future demand in a given region with the historical data, which can not estimate long-term consumption under different scenarios~\cite{xydas2016data}.

There are various avenues in which this work can be extended. Better understanding of the charging behavior of PEV drivers and the energy demand in residential regions would complete and enrich our planning estimates. As the PEVs are widely spreading these years, a stronger comprehension of the tie among mobility, socioeconomic characteristics of PEV owners, and the PEV incentive policies is necessary to accurately grasp the future energy demand as well as the pressure of the power grid. Another interesting avenue is to investigate PEVs management when the future energy structure changes, such as the rise of the wind and solar power.

%================================================
\section*{Materials and Methods}

%============
\paragraph*{Data}

The three main sources of data used in this study are described below.

\begin{enumerate}
    \item Mobile Phone Activity: Also referred to as Call Detail Records (CDRs), this data has been widely popular in the last decade, especially in the context of mobility modeling \cite{blondel2015survey,jiang2013review,alexander2014validation,
    toole2015path,colak2014analyzing,ccolak2016understanding}. For this work, we make use of the CDRs for the Bay Area including approximately 1.39 million users and more than 200 million calls they made over 6 weeks. Each record contains the anonymized user ID, timestamp, duration, and the geographic location of the associated cell tower. The spatial resolution is discretized to the service areas of 892 distinct cell towers. This information is used to build the TimeGeo mobility model for the Bay Area for a typical weekday. More details of the CDRs can be found in~\cite{ccolak2016understanding}
    
    \item PEV Charging Sessions: This data provided by Chargepoint Inc.~\cite{Chargepoint}, a charging station construction company, contains 580,000 records of PEV charging sessions in a part of commercial PEV supply equipment (EVSE) locations across the Bay Area in 2013, including any vehicle with a battery that can be charged. For each charging session, the following information is available: (i) one-time information on the EVSE location type, unique driver ID, total energy transferred, and plug-in/plug-out times; and (ii) charging power readings obtained every 15 minutes. The locations of the charging stations are anonymized to zipcode level. As a preprocessing step, we filter out records lasting shorter than 1 minute, are not in 2013, or have erroneous power measurements exceeding typical cable capacity and maximum charging rates.
    
    \item Census and Survey Information: The census data used in this study consists of shapefiles describing census tracts, their population, and income information~\cite{Census}. The survey information is obtained from the California Plug-in Electric Vehicle Driver Survey carried out in 2013 \cite{ev_survey}. This survey contains information on various sociodemographic characteristics and travel behavior of PEV drivers in California. We utilize information regarding income and average daily vehicle miles traveled in the estimation of PEV mobility.
\end{enumerate}

%============
\paragraph*{Individual Mobility Model\label{mobility}}

From the CDRs data, we are able to extract the visited places and time during the period of the dataset for each user. With that information, TimeGeo models and integrates the flexible temporal and spatial mobility choice of the individual. In the model, each day of a week is divided into 144 discrete intervals. For each interval, the individual decides to stay or move, and then where to go if she chooses to move. To represent the movement mechanisms, TimeGeo introduces time-inhomogeneous Markov chain model with three individual-specific mobility parameters: a weekly home-based tour number ($n_W$), a dwell rate ($\beta_1$), and a burst rate ($\beta_2$). Besides, $P(t)$ is defined as the global travel circadian rhythm of the population in an average week and it is different for commuters and non-commuters.

For the temporal movement choices, TimeGeo begins with determining if the individual is at home. If true, she will move with probability $n_w P(t)$, which represents her likelihood of making a trip originated from home in a time-interval $t$ of a week. If false, she will move with probability $\beta_1 n_w P(t)$. Then, if she decides to move, she goes to $other$ places with probability $\beta_2 n_w P(t)$ and goes back to home with probability $1-\beta_2 n_w P(t)$. The $P(t)$, distribution of $n_w$, $\beta_1 n_w$, and $\beta_2 n_w$ are illustrated in Supplementary Fig. 3. 

For the spatial movement choices, TimeGeo uses a rank-based exploration and preferential return (r-EPR) to determines the next place of the individual. In detail, when the individual chooses to move to an $other$ place, she could return to a visited place or explore a new place. The model assumes that the individual explores a new place with probability $P_{new} = \rho S^{-\gamma}$, which captures a decreasing propensity to visit new locations as the number of previously visited locations ($S$) increases with time. If the individual decides to return, the return location is selected from the visited locations according to her visiting frequency. If she decides to explore a new location, the alternative destinations are selected according to the distance to her origin with probability $P(k) \sim k^{-\alpha}$. More details of the TimeGeo model can be found in~\cite{jiang2016timegeo}.
To assess the simulation of individual mobility in Bay Area, we compare the aggregate performance of TimeGeo with NHTS and CHTS and show the results in Fig. 2, Supplementary Fig. 1 and Fig. 2.

%============
\paragraph*{Electric Vehicle Mobility Estimation\label{evOD}}

With the purpose of sampling PEV users from all vehicular drivers in Bay Area, we first extract the vehicular drivers from the entire population with the vehicle usage rate at census tract scale. Then, each vehicular driver is associated with a probability of using PEV, $P(EV|I_u, D_u)$ on the basis of the driver's household income $I_u$ and daily driving distance $D_u$. $I_{u}$ is the random variable that denotes the income of the trip maker and follows a standard normal distribution centered at the median income of the residential tract. The median income information at tract scale is from census data~\cite{Census}. $P(I_{u})$ is the probability density of the household income of all trip makers in Bay Area. Similarly, $D_{u}$ is the random variable that denotes the daily travel distance of the trip maker. The visited locations of the trip maker are obtained from the TimeGeo model and the routing distance is calculated by using a publicly available online API service for routing. $P(D_{u})$ is the probability density of the daily travel distance of all travelers in Bay Area. We assume that for a given trip maker, his or her income $I_{u}$ and daily travel distance $D_{u}$ are independent, thereby, $P(I_{u}, D_{u} \mid EV)=P(I_{u}\mid EV)P(D_{u}\mid EV)$, that is, $I_{u}$ and $D_{u}$ are also conditionally independent given a PEV driver.

To estimate the probability of using PEV, $P(EV|I_u,D_u)$, we begin by expressing the Bayesian relation,
\begin{equation}
P(EV \mid I_{u}, D_{u}) = \frac{P(I_{u}, D_{u} \mid EV ) P(EV)}{P(I_{u},D_{u})}
\label{eq:bayes}
\end{equation}
By imposing our aforementioned assumptions on Eq.~\ref{eq:bayes}, we have
\begin{equation}
P(EV \mid I_{u}, D_{u}) = \frac{P(I_{u}\mid EV) P(D_{u}\mid EV)P(EV)}{P(I_{u})P(D_{u})}
\label{eq:bayes2}
\end{equation}
In estimating this value, the share of PEVs within all cars in the Bay Area in 2013 equals to $0.62\%$ according to the CVRP data~\cite{CVRP}, that is, $P(EV)=0.62\%$. We make use of the PEV driver survey information regarding income and daily travel distance, namely $P(I_{u}\mid EV)$ and $P(D_{u}\mid EV)$, respectively. Once $P(EV \mid I_{u}, D_{u})$ is estimated, the probabilities are used to select the PEV drivers from all vehicular drivers. Fig. 3C represents the distribution of travel distance in the morning of all vehicular and PEV commuters.

%============

\paragraph*{Energy Consumption Models}
We design different energy consumption models for the four popular PEV modes in Bay Area. In detail, we estimate the power demand of Nissan Leaf with a drivetrain model and the trip information.
%\deleted{We infer the power demand of PEVs traveling into a selected zipcode via a drivetrain model that estimates the energy consumption of a trip given its length}~\cite{Saxena2015265}.
This drivetrain model builds the relationship between the energy consumption and two aggregate properties of the trip, the average travel speed and the route distance, which we estimate from a publicly available online API service for each PEV trip. That is,
\begin{equation}
E_{trip}^{Nissan} = f(V_{trip}) D_{trip}
\label{equ:drivetrain}
\end{equation}
where $V_{trip}$ and $D_{trip}$ are the average speed and route distance of the trip respectively. $f(V_{trip})$ implies the consumed power per mile (kWh/mi) when the PEV is traveling at speed $V_{trip}$ (mi/hr).
However, $f(V_{trip})$ depends on the battery used by the PEV model, meaning that different PEV models show different shapes of $f(V_{trip})$. In this work, we fit $f(V_{trip})$ with a piecewise linear function using the data observed from Nissan Leaf~\cite{needell2016potential}.
The curve of $f(V_{trip})$ is given in Supplementary Fig. 6B and the formulation is given as follows:
\begin{equation}
f(V_{trip}) = 
\begin{cases}
\begin{aligned}
& -23.12 \times 10^{-3} \cdot V_{trip} + 0.439 & \ \ V_{trip} \leq 6.70 \\
& -8.14 \times 10^{-3} \cdot V_{trip} + 0.338 & \ \ 6.70 \leq V_{trip} \leq 12.71 \\
& -0.38 \times 10^{-3} \cdot V_{trip} + 0.240 & \ \ 12.71 \leq V_{trip} \leq 21.75 \\
& 2.11 \times 10^{-3} \cdot V_{trip} + 0.185 & \ \ 21.75 \leq V_{trip} \leq 60.00 
\end{aligned}
\end{cases}
\label{equ:nissan}
\end{equation}

The consumption of Tesla Model S is the estimated by scaling the consumption of Nissan Leaf in the same trip by $1.229$,
as the Tesla model S consumes $22.9\%$ more energy on average than the Nissan Leaf~\cite{fiori2016power}. That is, $E_{trip}^{Tesla} = 1.229 f(V_{trip}) D_{trip}$.

For the PHEVs, we introduce the charge-depleting (CD) models to estimate their energy consumptions,
\begin{equation}
E_{trip}^{PHEV} = min \{D_{trip}\cdot r, C_{PHEV}\}
\end{equation}
where $r$ and $C_{PHEV}$ are the electricity consumption rate and the battery capacity of the PHEV, respectively. Wu et al. calibrated that $r=0.288$kWh/mi for PHEVs with 10 miles electric range; $r=0.337$kWh/mi for PHEVs with 20 miles electric range; and $r=0.342$kWh/mile for PHEVs with 40 miles electric range~\cite{wu2014cost}. In the Bay Area, the two most popular modes of PHEVs are Chevrolet Volt and Toyota Prius, and their electric ranges are 40 miles and 10 miles, respectively. The energy consumption models used here also match with the EV efficiency ratings released by the US Department of Energy (see Supplementary Table 1).

%\deleted{As shown in Supplementary Table 1, the average consumed energy per mile of Nissan Leaf is $0.30$ kWh/mi, and that of the second popular PEV, Chevrolet, is $0.29$ kWh/mi, which is quite similar to Nissan. Therefore, we argue that using the drivetrain model of Nissan is a good enough approximation in our analysis.}

%============
\paragraph*{Optimization Model\label{optim_formulation}}
We begin by discretizing a day into 15-minute intervals such that each day starts at $t=0$ and ends at $t=95$ \cite{Kara2015515}. For each charging session $i$ among $N$ in a day happen in a selected zipcode, we define $t^{i}_{a}$ as the arrival time index, $t^{i}_{c}$ as the time index where charging is complete, and $t^{i}_{d}$ as the departure time index. We represent the time indices by the vector $\tau^{i}$, and the power consumption by vectors $\bm{P}^{i}$ and $\bm{Q}^{i}$, all defined as follows:
\begin{equation}
\begin{aligned}
\bm{\tau}^{i} &= [t^{i}_{a}, \hdots, t^{i}_{c}]^\intercal \\
\bm{P}^{i} &= [P^{i}_{0}, \hdots, P^{i}_{95}]^\intercal \\
\bm{Q}^{i} &= [P^{i}_{t^{i}_{a}} , \hdots , P^{i}_{t^{i}_{c}}]^\intercal \\
\end{aligned}
 \end{equation} 

By shifting $\bm{Q}^{i}$ within $\bm{P}^{i}$ by an amount $d^{i}$ for all sessions, we can modify the overall power demand curve. We define $M^{i}= (t^{i}_{c} - t^{i}_{a}) +1 $ as the total number of non-zero power measurements in this charging session (i.e. total number of elements in $\bm{Q}^{i}$), given that charging sessions start immediately upon arrival. We enforce continuity of the charging process, the non-violation of departure times, and amounts of session energy.

To capture the constraints proposed above, we introduce the following formal constraints:

\begin{equation}
\label{eq:constraints}
\left.
\begin{aligned}
& \tau^{i}_{j} \geq 0\\
& \tau^{i}_{j} \leq 95\\
& \tau^{i}_{j} \geq t^{i}_{a}+d^{i}\\
& \tau^{i}_{j} \leq t^{i}_{c}+d^{i}\\
& t^{i}_{d} \geq t^{i}_{c}+d^{i}\\
& \tau^{i}_{j} < \tau^{i}_{j+1}\\
\end{aligned}
\right\} 
\begin{aligned} &\forall i \in [1,N],\\ &\forall j \in [1,M^{i}] \end{aligned} 
\end{equation}
where $d^i$ is the delay of the charging session of the $i$-th PEV driver. As the mobility motif of the PEV driver limits the acceptability of the recommendations, we customize $d^i$ for the PEV drivers with different mobility motifs as shown in Fig. 5B. Assuming that the delay of strategy is $d$, we introduce the following constraints:
\begin{equation}
d^i = 
\begin{cases}
\begin{aligned}
& d & & \text{H-W-H}\\
& min\{0, d\} & & \text{H-W-O-H}\\
& 0 & & \text{H-O-W-H \ \& H-O-W-O-H}
%& 0 & & \text{H-O-W-O-H}
\end{aligned}
\end{cases}
\end{equation}
In this customization of delay, the drivers with mobility motif \emph{home-work-home} (H-W-H) could accept any change of arrival and departure time; the drivers with mobility motif \emph{home-work-other-home} (H-W-O-H) can not delay their departure time, that is, the delay $d$ must be non-positive; the drivers with mobility motif \emph{home-other-work-home} (H-O-W-H) can not change their arrive time; the drivers with mobility motif \emph{home-other-work-other-home} (H-O-W-O-H) can change neither their arrival time nor departure time.

We construct the proposed constraints using a binary decision matrix to represent charging or non-charging time slots within the optimization duration. To represent the candidate time slot at which $Q^i_j$ can be positioned, we create binary row vectors $\bm{x^{i}_{j}}$ each consisting of 95 binary decision variables: $x^{i}_{j,k} \in \{0,1\}, \forall j \in [1,M^{i}], \forall i \in[1,N], \forall k \in [0,95]$.
\begin{equation}
\bm{X^{i}}= \begin{bmatrix} \bm{x^{i}_{1}} \\ \vdots\\ \bm{x^{i}_{M^{i}}}\end{bmatrix} = \begin{bmatrix} x^{i}_{1,0} & \hdots & x^{i}_{1,95} \\
 \vdots& \ddots & \vdots \\
 x^{i}_{M^{i},0} & \hdots & x^{i}_{M^{i},95} 
 \end{bmatrix}
 \end{equation}
 
Finally, we write the variables in the constraints given in~(\ref{eq:constraints}) using the binary decision variable as follows:

\begin{equation}
\label{eq:optim}
\begin{aligned}
& \bm{\tau^{i}}=\bm{X^{i}}\begin{bmatrix} 0\\ \vdots\\ 95\end{bmatrix} \end{aligned} 
\end{equation}

The aggregate power vector $\bm{AP}$ is given as follows:

\begin{equation}
\bm{AP} = \sum_{i=0}^N \bm{P}^{i}= \begin{bmatrix} \bm{Q}^{1}\\ \vdots \\ \bm{Q}^{N} \end{bmatrix}^\intercal \begin{bmatrix} \bm{X}^{1}\\ \vdots \\ \bm{X}^{N}\end{bmatrix}\\
\end{equation}

The resulting formulation is a mixed-integer linear program, with decision variables $\bm{X}^{i}$, $P_{peak}$, and $d^{i}$ of which the latter two are integers. The problem can be proposed to minimize the daily peak load $P_{peak}$ for a group of PEVs arriving to the same zip code location:

\begin{equation*}
\begin{aligned}
& \underset{\bm{X}^{i}, P_{peak},d^{i}}{\text{minimize}}
& P_{peak} \\
\end{aligned}
\end{equation*}

subject to~(\ref{eq:constraints}) and the following additional constraints:

\begin{equation}
AP^{i}_{t} \le P_{peak}, \hskip5pt \forall i \in [1,N], \hskip5pt \forall t \in [0,95]
\end{equation}

% Your references go at the end of the main text, and before the
% figures.  For this document we've used BibTeX, the .bib file
% scibib.bib, and the .bst file Science.bst.  The package scicite.sty
% was included to format the reference numbers according to *Science*
% style.

\bibliography{sciadvbib}

\begin{thebibliography}{10}

\bibitem{michalek2011valuation}
J.~J. Michalek, {\it et~al.\/}, Valuation of plug-in vehicle life-cycle air
  emissions and oil displacement benefits.
\newblock {\it Proceedings of the National Academy of Sciences\/} {\bf 108},
  16554--16558 (2011).

\bibitem{atia2015more}
R.~Atia, N.~Yamada, More accurate sizing of renewable energy sources under high
  levels of electric vehicle integration.
\newblock {\it Renewable Energy\/} {\bf 81}, 918--925 (2015).

\bibitem{needell2016potential}
Z.~A. Needell, J.~McNerney, M.~T. Chang, J.~E. Trancik, Potential for
  widespread electrification of personal vehicle travel in the united states.
\newblock {\it Nature Energy\/} {\bf 1}, 16112 (2016).

\bibitem{nykvist2015rapidly}
B.~Nykvist, M.~Nilsson, Rapidly falling costs of battery packs for electric
  vehicles.
\newblock {\it Nature Climate Change\/} {\bf 5}, 329--332 (2015).

\bibitem{melton2016moving}
N.~Melton, J.~Axsen, D.~Sperling, Moving beyond alternative fuel hype to
  decarbonize transportation.
\newblock {\it Nature Energy\/} {\bf 1}, 16013 (2016).

\bibitem{hu2016integrated}
X.~Hu, S.~J. Moura, N.~Murgovski, B.~Egardt, D.~Cao, Integrated optimization of
  battery sizing, charging, and power management in plug-in hybrid electric
  vehicles.
\newblock {\it IEEE Transactions on Control Systems Technology\/} {\bf 24},
  1036--1043 (2016).

\bibitem{deshazo2016improving}
J.~DeShazo, Improving incentives for clean vehicle purchases in the united
  states: Challenges and opportunities.
\newblock {\it Review of Environmental Economics and Policy\/} {\bf 10},
  149--165 (2016).

\bibitem{ev_outlook}
T.~Trigg, {\it et~al.\/}, Global ev outlook: understanding the electric vehicle
  landscape to 2020.
\newblock {\it Int. Energy Agency\/} pp. 1--40 (2013).

\bibitem{hines2009large}
P.~Hines, J.~Apt, S.~Talukdar, Large blackouts in north america: Historical
  trends and policy implications.
\newblock {\it Energy Policy\/} {\bf 37}, 5249--5259 (2009).

\bibitem{brummitt2013transdisciplinary}
C.~D. Brummitt, P.~D. Hines, I.~Dobson, C.~Moore, R.~M. D'Souza,
  Transdisciplinary electric power grid science.
\newblock {\it Proceedings of the National Academy of Sciences\/} {\bf 110},
  12159--12159 (2013).

\bibitem{buldyrev2010catastrophic}
S.~V. Buldyrev, R.~Parshani, G.~Paul, H.~E. Stanley, S.~Havlin, Catastrophic
  cascade of failures in interdependent networks.
\newblock {\it Nature\/} {\bf 464}, 1025--1028 (2010).

\bibitem{brummitt2012suppressing}
C.~D. Brummitt, R.~M. D’Souza, E.~Leicht, Suppressing cascades of load in
  interdependent networks.
\newblock {\it Proceedings of the National Academy of Sciences\/} {\bf 109},
  E680--E689 (2012).

\bibitem{pahwa2014abruptness}
S.~Pahwa, C.~Scoglio, A.~Scala, Abruptness of cascade failures in power grids.
\newblock {\it Scientific reports\/} {\bf 4} (2014).

\bibitem{McAndrew2015}
T.~C. McAndrew, C.~M. Danforth, J.~P. Bagrow, Robustness of spatial
  micronetworks.
\newblock {\it Phys. Rev. E\/} {\bf 91}, 042813 (2015).

\bibitem{mwasilu2014electric}
F.~Mwasilu, J.~J. Justo, E.-K. Kim, T.~D. Do, J.-W. Jung, Electric vehicles and
  smart grid interaction: A review on vehicle to grid and renewable energy
  sources integration.
\newblock {\it Renewable and Sustainable Energy Reviews\/} {\bf 34}, 501--516
  (2014).

\bibitem{halu2016data}
A.~Halu, A.~Scala, A.~Khiyami, M.~C. Gonz{\'a}lez, Data-driven modeling of
  solar-powered urban microgrids.
\newblock {\it Science Advances\/} {\bf 2}, e1500700 (2016).

\bibitem{mureddu2015green}
M.~Mureddu, G.~Caldarelli, A.~Chessa, A.~Scala, A.~Damiano, Green power grids:
  How energy from renewable sources affects networks and markets.
\newblock {\it PloS one\/} {\bf 10}, e0135312 (2015).

\bibitem{bayram2012smart}
I.~S. Bayram, G.~Michailidis, M.~Devetsikiotis, F.~Granelli, S.~Bhattacharya,
  {\it Control and Optimization Methods for Electric Smart Grids\/} (Springer,
  2012), pp. 133--145.

\bibitem{callaway2011achieving}
D.~S. Callaway, I.~A. Hiskens, Achieving controllability of electric loads.
\newblock {\it Proceedings of the IEEE\/} {\bf 99}, 184--199 (2011).

\bibitem{moura2011stochastic}
S.~J. Moura, H.~K. Fathy, D.~S. Callaway, J.~L. Stein, A stochastic optimal
  control approach for power management in plug-in hybrid electric vehicles.
\newblock {\it IEEE Transactions on control systems technology\/} {\bf 19},
  545--555 (2011).

\bibitem{clement2010impact}
K.~Clement-Nyns, E.~Haesen, J.~Driesen, The impact of charging plug-in hybrid
  electric vehicles on a residential distribution grid.
\newblock {\it Power Systems, IEEE Transactions on\/} {\bf 25}, 371--380
  (2010).

\bibitem{tal2014charging}
G.~Tal, M.~Nicholas, J.~Davies, J.~Woodjack, Charging behavior impacts on
  electric vehicle miles traveled: Who is not plugging in?
\newblock {\it Transportation Research Record: Journal of the Transportation
  Research Board\/} pp. 53--60 (2014).

\bibitem{harris2014empirically}
C.~B. Harris, M.~E. Webber, An empirically-validated methodology to simulate
  electricity demand for electric vehicle charging.
\newblock {\it Applied Energy\/} {\bf 126}, 172--181 (2014).

\bibitem{lin2014optimizing}
Z.~Lin, Optimizing and diversifying electric vehicle driving range for us
  drivers.
\newblock {\it Transportation Science\/} {\bf 48}, 635--650 (2014).

\bibitem{rajakaruna2015plug}
S.~Rajakaruna, F.~Shahnia, A.~Ghosh, {\it Plug In Electric Vehicles in Smart
  Grids\/} (Springer, 2015).

\bibitem{tamor2015rapid}
M.~A. Tamor, P.~E. Moraal, B.~Reprogle, M.~Mila{\v{c}}i{\'c}, Rapid estimation
  of electric vehicle acceptance using a general description of driving
  patterns.
\newblock {\it Transportation Research Part C: Emerging Technologies\/} {\bf
  51}, 136--148 (2015).

\bibitem{hines2014understanding}
P.~Hines, {\it et~al.\/}, Understanding and managing the impacts of electric
  vehicles on electric power distribution systems, {\it Tech. rep.\/},
  University of Vermont (2014).

\bibitem{yuksel2015effects}
T.~Yuksel, J.~J. Michalek, Effects of regional temperature on electric vehicle
  efficiency, range, and emissions in the united states.
\newblock {\it Environmental science \& technology\/} {\bf 49}, 3974--3980
  (2015).

\bibitem{rezaei2014packetized}
P.~Rezaei, J.~Frolik, P.~D. Hines, Packetized plug-in electric vehicle charge
  management.
\newblock {\it Smart Grid, IEEE Transactions on\/} {\bf 5}, 642--650 (2014).

\bibitem{valogianni2014effective}
K.~Valogianni, W.~Ketter, J.~Collins, D.~Zhdanov, {\it AAAI\/} (2014), pp.
  472--478.

\bibitem{ma2013decentralized}
Z.~Ma, D.~S. Callaway, I.~A. Hiskens, Decentralized charging control of large
  populations of plug-in electric vehicles.
\newblock {\it Control Systems Technology, IEEE Transactions on\/} {\bf 21},
  67--78 (2013).

\bibitem{Kara2015515}
E.~C. Kara, {\it et~al.\/}, Estimating the benefits of electric vehicle smart
  charging at non-residential locations: A data-driven approach.
\newblock {\it Applied Energy\/} {\bf 155}, 515 - 525 (2015).

\bibitem{subramanian2013real}
A.~Subramanian, M.~J. Garcia, D.~S. Callaway, K.~Poolla, P.~Varaiya, Real-time
  scheduling of distributed resources.
\newblock {\it Smart Grid, IEEE Transactions on\/} {\bf 4}, 2122--2130 (2013).

\bibitem{yang2014risk}
L.~Yang, J.~Zhang, H.~V. Poor, Risk-aware day-ahead scheduling and real-time
  dispatch for electric vehicle charging.
\newblock {\it Smart Grid, IEEE Transactions on\/} {\bf 5}, 693--702 (2014).

\bibitem{zakariazadeh2014multi}
A.~Zakariazadeh, S.~Jadid, P.~Siano, Multi-objective scheduling of electric
  vehicles in smart distribution system.
\newblock {\it Energy Conversion and Management\/} {\bf 79}, 43--53 (2014).

\bibitem{garcia2014plug}
J.~Garc{\'\i}a-Villalobos, I.~Zamora, J.~San~Mart{\'\i}n, F.~Asensio,
  V.~Aperribay, Plug-in electric vehicles in electric distribution networks: A
  review of smart charging approaches.
\newblock {\it Renewable and Sustainable Energy Reviews\/} {\bf 38}, 717--731
  (2014).

\bibitem{alizadeh2015joint}
M.~Alizadeh, {\it et~al.\/}, Joint management of electric vehicles in coupled
  power and transportation networks.
\newblock {\it arXiv preprint arXiv:1511.03611\/}  (2015).

\bibitem{jiang2016timegeo}
S.~Jiang, {\it et~al.\/}, The timegeo modeling framework for urban mobility
  without travel surveys.
\newblock {\it Proceedings of the National Academy of Sciences\/} p. 201524261
  (2016).

\bibitem{jiang2013review}
S.~Jiang, {\it et~al.\/}, {\it Proceedings of the 2nd ACM SIGKDD International
  Workshop on Urban Computing\/} (ACM, 2013), p.~2.

\bibitem{colak2014analyzing}
S.~{\c{C}}olak, L.~P. Alexander, B.~G. Alvim, S.~R. Mehndiratta, M.~C.
  Gonz{\'a}lez, Analyzing cell phone location data for urban travel: current
  methods, limitations, and opportunities.
\newblock {\it Transportation Research Record: Journal of the Transportation
  Research Board\/} pp. 126--135 (2015).

\bibitem{toole2015path}
J.~L. Toole, {\it et~al.\/}, The path most traveled: Travel demand estimation
  using big data resources.
\newblock {\it Transportation Research Part C: Emerging Technologies\/} {\bf
  58}, 162 - 177 (2015).

\bibitem{CHTS2010}
{National Renewable Energy Laboratory}, Transportation secure data center,
  www.nrel.gov/tsdc (2015). [Online; accessed 15-Jan-2015].

\bibitem{NHTS2009}
{U.S. Department of Transportation, Federal Highway Administration}, National
  household travel survey, http://nhts.ornl.gov (2009). [Online; accessed
  01-Oct-2016].

\bibitem{schneider2013unravelling}
C.~M. Schneider, V.~Belik, T.~Couronn{\'e}, Z.~Smoreda, M.~C. Gonz{\'a}lez,
  Unravelling daily human mobility motifs.
\newblock {\it Journal of The Royal Society Interface\/} {\bf 10}, 20130246
  (2013).

\bibitem{ev_survey}
{California Center for Sustainable Energy}, California plug-in electric vehicle
  driver survey results - may 2013, {\it Tech. rep.\/}, California Center for
  Sustainable Energy (2013).

\bibitem{CVRP}
{Center for Sustainable Energy}, California air resources board clean vehicle
  rebate project, rebate statistics,
  https://cleanvehiclerebate.org/rebate-statistics (2017). [Online; accessed
  05-April-2017].

\bibitem{Commute2017}
{Vital Signs}, Commute time, http://www.vitalsigns.mtc.ca.gov/commute-time
  (2017). [Online; accessed 16-May-2017].

\bibitem{Saxena2015265}
S.~Saxena, C.~L. Floch, J.~MacDonald, S.~Moura, Quantifying {EV} battery
  end-of-life through analysis of travel needs with vehicle powertrain models.
\newblock {\it Journal of Power Sources\/} {\bf 282}, 265 - 276 (2015).

\bibitem{wu2014cost}
X.~Wu, J.~Dong, Z.~Lin, Cost analysis of plug-in hybrid electric vehicles using
  gps-based longitudinal travel data.
\newblock {\it Energy Policy\/} {\bf 68}, 206--217 (2014).

\bibitem{yilmaz}
M.~Yilmaz, P.~T. Krein, Review of battery charger topologies, charging power
  levels, and infrastructure for plug-in electric and hybrid vehicles.
\newblock {\it IEEE Transactions on Power Electronics\/} {\bf 28}, 2151-2169
  (2013).

\bibitem{oleksek}
{U.S. Department of Energy}, Workplace charging challenge, mid-program review:
  Employees plug in, {\it Tech. rep.\/}, {U.S. Department of Energy} (2015).

\bibitem{billing}
{Pacific Gas and Electric Company}, Electric schedule e-19: Medium general
  demand-metered {TOU} service (2010).

\bibitem{merugu2009incentive}
D.~Merugu, B.~S. Prabhakar, N.~Rama, {\it Proc. of ACM NetEcon Workshop\/}
  (Citeseer, 2009).

\bibitem{Xu2017}
S.~Xu, E.~Barbour, M.~C. Gonz{\'a}lez, {\it Proceedings of the 6nd ACM SIGKDD
  International Workshop on Urban Computing\/} (ACM, 2017), p.~2.

\bibitem{luo2017electric}
X.~Luo, T.~Hong, Y.~Chen, M.~A. Piette, Electric load shape benchmarking for
  small-and medium-sized commercial buildings.
\newblock {\it Applied Energy\/} {\bf 204}, 715--725 (2017).

\bibitem{xydas2016data}
E.~Xydas, {\it et~al.\/}, A data-driven approach for characterising the
  charging demand of electric vehicles: A uk case study.
\newblock {\it Applied Energy\/} {\bf 162}, 763--771 (2016).

\bibitem{blondel2015survey}
V.~D. Blondel, A.~Decuyper, G.~Krings, A survey of results on mobile phone
  datasets analysis.
\newblock {\it EPJ Data Science\/} {\bf 4:10} (2015).

\bibitem{alexander2014validation}
L.~Alexander, S.~Jiang, M.~Murga, M.~C. Gonz{\'a}lez, Origin–destination
  trips by purpose and time of day inferred from mobile phone data.
\newblock {\it Transportation Research Part C: Emerging Technologies\/} {\bf
  58, Part B}, 240 - 250 (2015).

\bibitem{ccolak2016understanding}
S.~{\c{C}}olak, A.~Lima, M.~C. Gonz{\'a}lez, Understanding congested travel in
  urban areas.
\newblock {\it Nature communications\/} {\bf 7} (2016).

\bibitem{Chargepoint}
{Chargepoint Inc.}, {https://www.chargepoint.com/} (2016).

\bibitem{Census}
{United States Census Bureau}, Census data, https://www.census.gov/data.html
  (2016). [Online; accessed 15-Oct-2016].

\bibitem{fiori2016power}
C.~Fiori, K.~Ahn, H.~A. Rakha, Power-based electric vehicle energy consumption
  model: Model development and validation.
\newblock {\it Applied Energy\/} {\bf 168}, 257--268 (2016).

\end{thebibliography}
\bibliographystyle{ScienceAdvances}

\noindent \textbf{Acknowledgements:} 
We would like to thank Chargepoint Inc. for providing the electric vehicle charging data, and Airsage for providing the call detail records used in this study. We also would like to thank Sila Kiliccote and Michaelangelo Tabone for their valuable feedback.\\
\noindent \textbf{Funding:} This work was supported by the Siebel Institute and MIT Energy Initiative.\\
\noindent \textbf{Author Contributions} YX, SC, and ECK conceived the research and designed the analyses. YX, SC, and MCG performed the analyses and wrote the paper. ECK helped perform the analyses. SJM and MCG provided general advice and supervised the research.\\
\noindent \textbf{Competing Interests} The authors declare that they have no competing financial interests.\\
\noindent \textbf{Data and materials availability:} All data needed to evaluate the conclusions in the paper are present in the paper. Additional data related to this paper may be requested from the authors.

\clearpage
\begin{figure*}
\centerline{\includegraphics[width=0.9\linewidth]{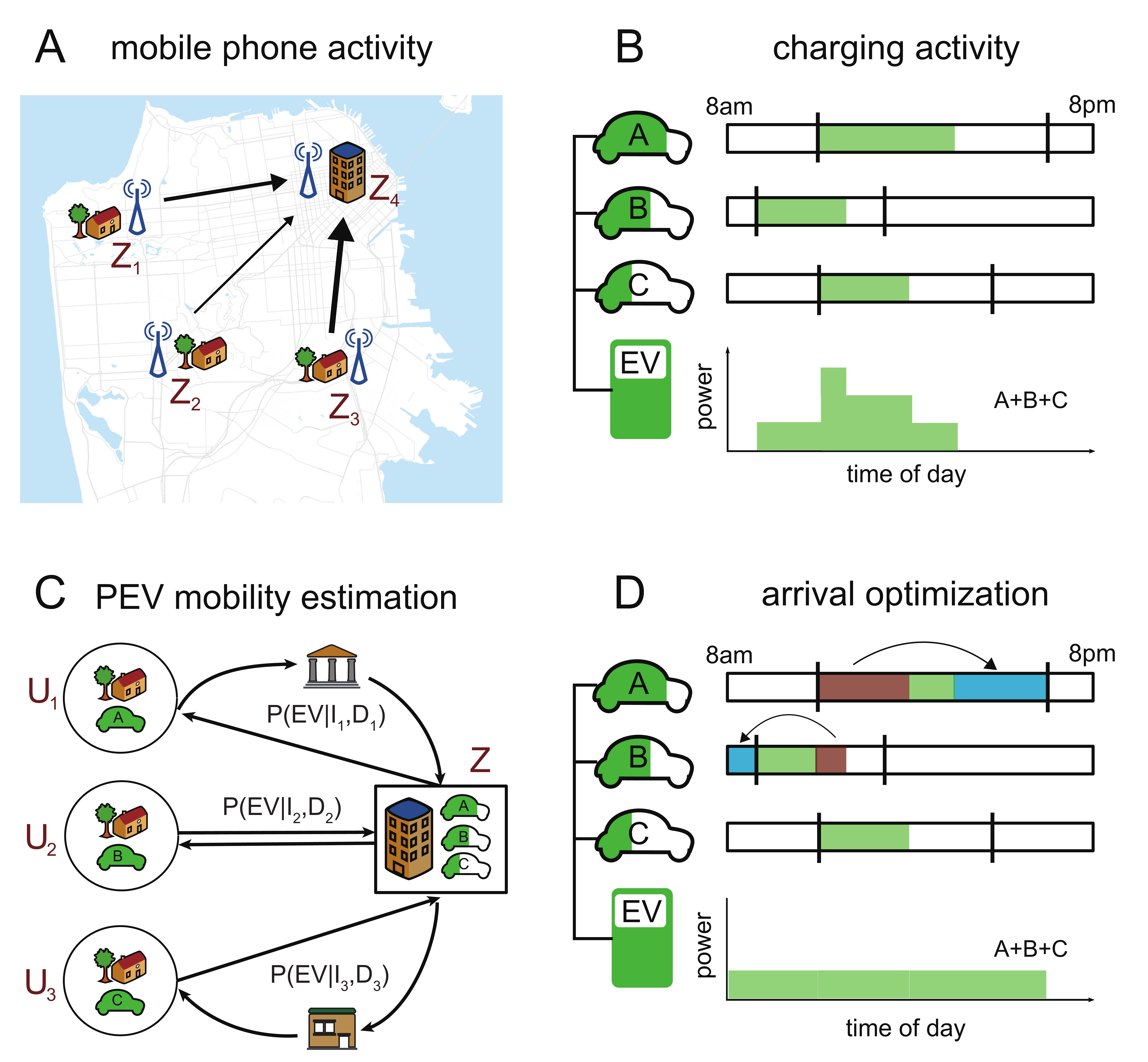}}
\end{figure*}
\noindent {\bf Fig. 1.} \textbf{Coupling PEV charging with urban mobility.} \textbf{A} Mobile phone data is used to model individual mobility. \textbf{B} Charging sessions data is used to characterize individual and total electricity demand curves. \textbf{C} The Bayesian inference method is proposed to find the probability that the vehicular trip is made by a PEV. Z is the work zipcode of users $U_1$, $U_2$, and $U_3$. \textbf{D} Charging activity is shifted to create a recommendation scheme that relieve peaks in demand and generate monetary savings to the drivers based on electricity tariffs.

\clearpage
\noindent \textbf{Table 1.} \textbf{Characteristics of PEV drivers.} Distribution of \textbf{A} average daily miles driven and \textbf{B} annual income by PEV drivers in California, USA \cite{ev_survey}.
\begin{table}[!h]
\begin{center}
\label{tab:ev_survey}
\begin{tabular}{c@{\hskip 1cm}c}%
\begin{tabular}{rcc}
(1000\$) & \textbf{Conventional} & \textbf{PEV} \\
\hline
Unknown & 20\% & 17\% \\
$<$ 50 & 20\% & 2\% \\
50-100 & 30\% & 13\% \\
100-150 & 14\% & 20\% \\
$>$ 250 & 15\% & 47\% \\
\hline
\end{tabular}
\hspace{2cm}
\begin{tabular}{rc}
(miles) & \textbf{\%} \\
\hline
$<$ 15 & 14\% \\
15-30 & 50\% \\
30-45 & 28\% \\
$>$ 45 & 8\% \\
\hline
\end{tabular}

\end{tabular}
\end{center}
\end{table}

\clearpage
\begin{figure*}
\centerline{\includegraphics[width=\linewidth]{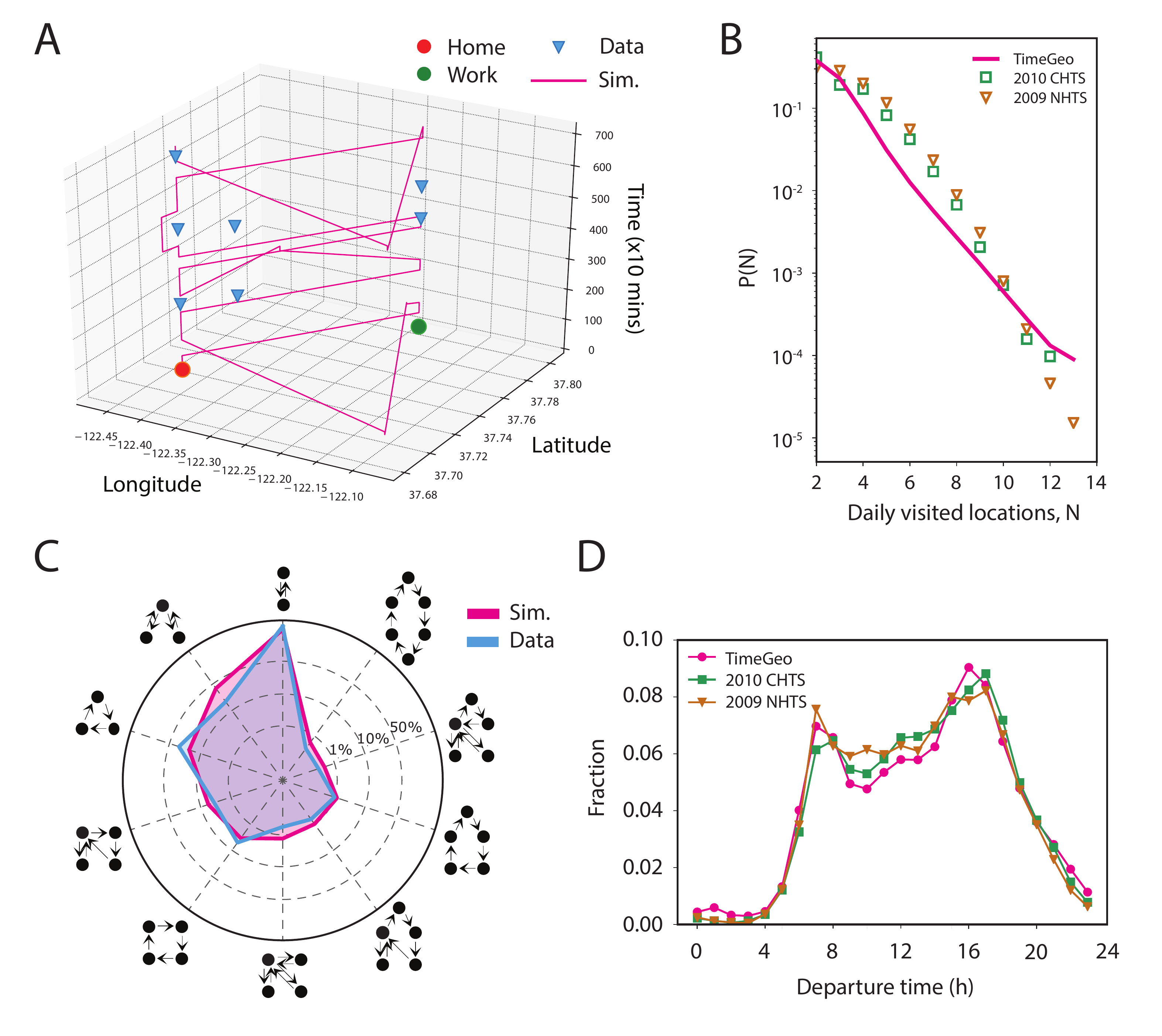}}
\end{figure*}
\noindent {\bf Fig. 2.} \textbf{Validation of individual mobility simulation in Bay Area.} \textbf{A} Simulated trajectory of a mobile phone user who is labeled as commuter. The blue triangles represent the actually recorded activities of the selected user from October 22 to 25, 2012. The red and green circles are the recognized home and work from the mobile phone data respectively. \textbf{B} Population distribution comparison of daily visited locations between simulation and two travel survey datasets, 2010 CHTS, and 2009 NHTS. Supplementary Fig. 1A and 1B respectively present the comparisons of two more properties of mobility, stay duration and trip distance. \textbf{C} Validation of individual mobility motifs for the active users in simulation and CDRs data. The distribution of the 10 primary motifs shows a high degree of similarity to the CDRs data. The comparison of motifs of commuters and non-commuters are shown in the Supplementary Fig. 1C. \textbf{D} Fraction of trip departures by time of the day, comparing the simulation, the 2010 CHTS, and the 2009 NHTS. Supplementary Fig. 2 shows the comparisons for various trip purposes.

\clearpage
\begin{figure*}
\centerline{\includegraphics[width=0.98\linewidth]{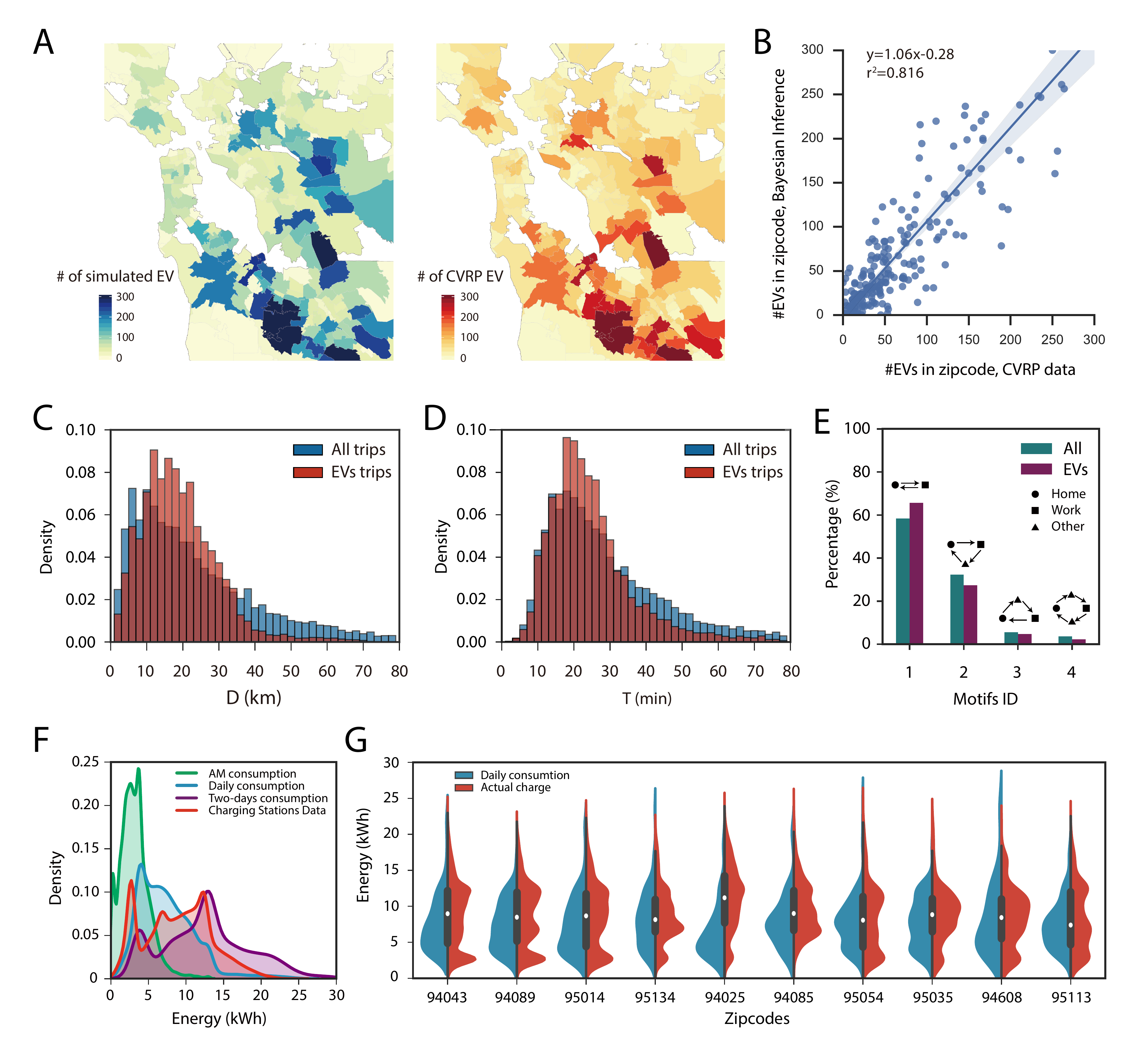}}
\end{figure*}
\noindent {\bf Fig. 3.} \textbf{Validation of PEV mobility estimation and calibration of PEV charging behavior.} \textbf{A} Number of PEVs in the residential zipcodes of the Bay Area from the simulation and the CVRP datasets in the end of 2013. The total number of PEVs is $15963$ from our simulation, which is close to the actual number from CVRP datasets, $16103$. \textbf{B} Correlation between the simulated PEV and the PEV from CVRP data. \textbf{C, D} Probability distributions of commuting distances, $D$, and commuting travel times, $T$, of all vehicle trips and EV trips estimated through income information and trip distances. \textbf{E} Fractions of four types of mobility motifs of all commuters and PEV users. $58\%$ of commuters only travel between home and work on weekdays, while the rest $42\%$ have other activities before or after work. Similarly, $66\%$ of commuters using PEVs only travel between home and work, and the rest $34\%$ have other activities before or after work. \textbf{F} Probability distributions of charging energy $E_{S}$ obtained from charging sessions compared to those of the energy demand estimated by energy consumption model on three charging behavior scenarios, morning, daily, and two-days. \textbf{G} Probability distributions of charging energy $E_{S}$ and those of the daily energy demand of simulated PEVs for the zipcodes which have most charging session records.

\clearpage
\begin{figure*}
\centerline{\includegraphics[width=\linewidth]{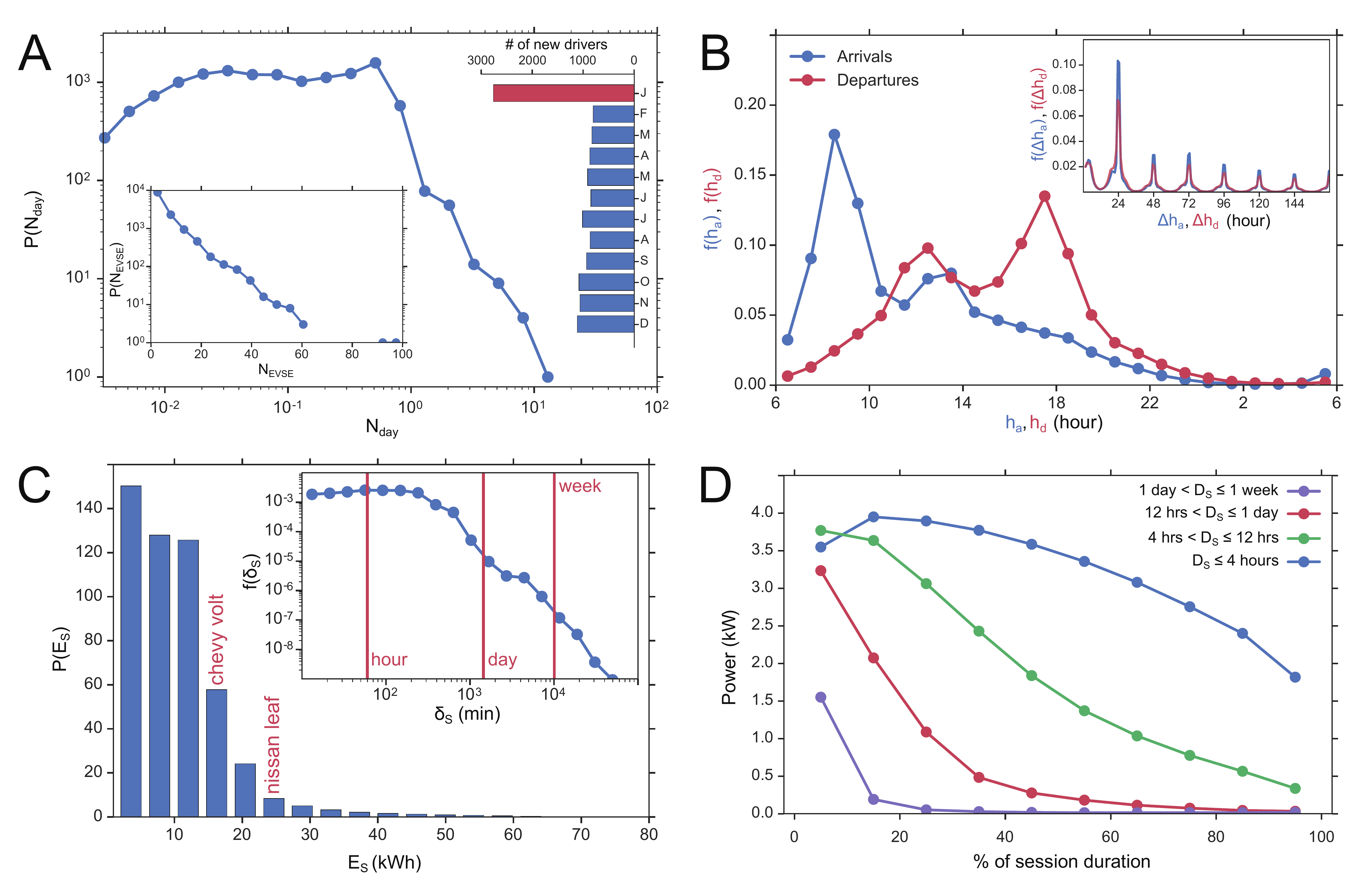}}
\end{figure*}
\noindent {\bf Fig. 4.} \textbf{PEV charging session profiles.} \textbf{A} Distribution of $N_{day}$ the number of sessions per day for each driver ID starting from the day of first record (right inset: number of new driver IDs added every month, left inset: distribution of $N_{EVSE}$, the number of unique EVSEs visited by every driver ID.) \textbf{B} Distributions of $h_a$ and $h_d$, the arrival and departure hours to and from an EVSE. (inset: the distributions of $\Delta h_{a}$ and $\Delta h_{d}$, the interarrival and interdeparture times for a driver ID visiting a specific EVSE.) \textbf{C} Distribution of $E_S$, the total energy withdrawn per session (inset: the distribution of $\delta_S$, session durations.) \textbf{D} Power consumption as a function of the normalized session duration segmented by total duration groups.

\clearpage
\begin{figure*}
\centerline{\includegraphics[width=0.92\linewidth]{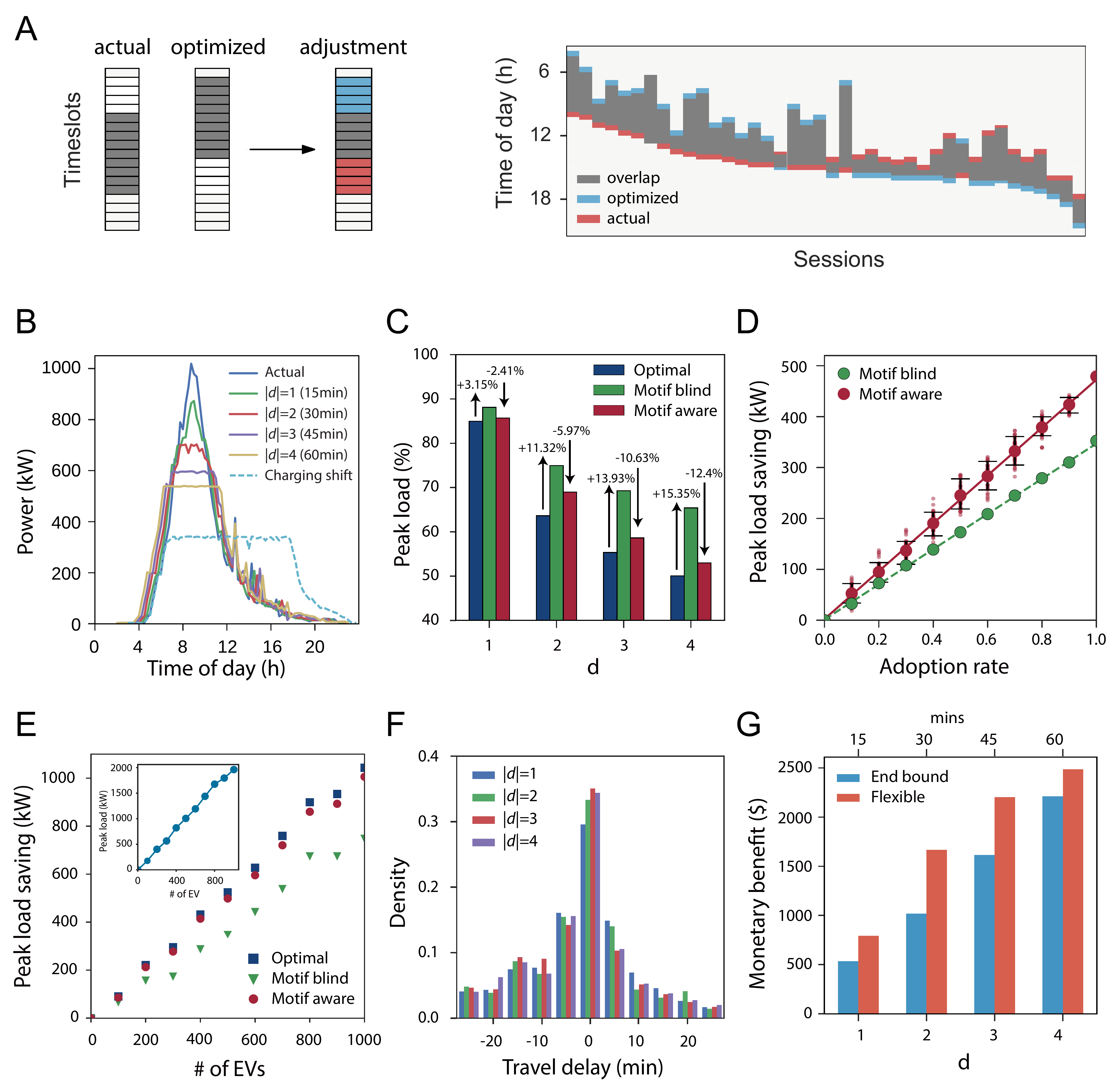}}
\label{fig:timeshift}
\end{figure*}
 {\bf Fig. 5.} \textbf{Assessing the benefits of minimizing peak power.} \textbf{A} An illustration of a sample of sessions in the \emph{flexible} strategy. The charging event within a session is shifted such that the overall peak power is minimized. \textbf{B} Decrease in peak power measurements for the varying $d$ of the \emph{flexible} and \emph{charging-shift} strategy. The flexible time shifts are constrained by mobility motifs of PEV users, and the peak shaving of $47\%$ can be achieved. The \emph{charging-shift} strategy reduces the peak load by $66\%$. \textbf{C} Percentage peak load of the three schemes \emph{Optimal}, \emph{Motif blind} and \emph{Motif aware} are presented. The positive value on the bar implies the gap in the estimates between the \emph{Optimal} vs. \emph{Motif blind}, showing the effects of lacking mobility information. The negative value on the bars implies the improvements using \emph{Motif aware} information vs. \emph{Motif blind}, customizing the optimization strategy to the mobility constraints. \textbf{D} Peak load saving versus PEV driver's adoption rate of the time-shift recommendations, for the \emph{Motif-blind} and \emph{Motif-aware} schemes when $d=4$. For each adoption rate, we randomly choose PEV drivers who accept the recommendations and average over $50$ realizations. The error bar shows the $10th$ and $90th$ percentile of the savings of \emph{Motif aware} for the given acceptance rate. The red solid (green dash) line presents the linear fit between the saving of \emph{Morif aware} (\emph{blind}) and adoption rate.
\textbf{E} Projected peak load saving of the \emph{flexible} strategy when $d=4$ vs. the number of PEV users working in the selected zipcode. Inset shows the projected peak load versus the number of PEV users.
\textbf{F} Changes on travel times of morning and evening trips. The majority of the drivers are not influenced, the worst case is a few individuals suffering from an additional 20 minute delays for $d=4$. \textbf{G} Monthly savings for varying values of $d$. The \emph{flexible} strategy restrained by mobility motifs is able to generate savings worth approximately 2500\$ for $d=4$.

\end{document}